\documentclass[reprint,amsmath,amssymb,apr,aip,onecolumn, 10pt]{revtex4-1}
\usepackage{graphicx}
\usepackage{graphics}
\usepackage{mathptmx}
\usepackage{times}
\usepackage{amsmath}
\usepackage{amssymb}
\usepackage{dcolumn}
\usepackage{bm}
\usepackage{units}
\usepackage{color}
\usepackage{soul}
\usepackage{multirow}
\newcommand{\floor}[1]{\lfloor #1 \rfloor}

\graphicspath{{Figures/}}

\begin{document}
\title{Mixed-precision deep learning based on computational memory}
\author{S. R. Nandakumar}\affiliation{IBM Research - Zurich, 8803 R\"{u}schlikon, Switzerland}\affiliation{New Jersey Institute of Technology (NJIT), Newark, NJ 07102, USA}
\author{Manuel Le Gallo}\email{anu@zurich.ibm.com}\affiliation{IBM Research - Zurich, 8803 R\"{u}schlikon, Switzerland}
\author{Christophe Piveteau}\affiliation{IBM Research - Zurich, 8803 R\"{u}schlikon, Switzerland}\affiliation{ETH Zurich, 8092 Zurich, Switzerland}
\author{Vinay Joshi}\affiliation{IBM Research - Zurich, 8803 R\"{u}schlikon, Switzerland}\affiliation{New Jersey Institute of Technology (NJIT), Newark, NJ 07102, USA}
\author{Giovanni Mariani}\affiliation{IBM Research - Zurich, 8803 R\"{u}schlikon, Switzerland}
\author{Irem Boybat}\affiliation{IBM Research - Zurich, 8803 R\"{u}schlikon, Switzerland}\affiliation{Ecole Polytechnique Federale de Lausanne (EPFL), 1015 Lausanne, Switzerland}
\author{Geethan Karunaratne}\affiliation{IBM Research - Zurich, 8803 R\"{u}schlikon, Switzerland}\affiliation{ETH Zurich, 8092 Zurich, Switzerland}
\author{Riduan Khaddam-Aljameh}\affiliation{IBM Research - Zurich, 8803 R\"{u}schlikon, Switzerland}\affiliation{ETH Zurich, 8092 Zurich, Switzerland}
\author{Urs Egger}\affiliation{IBM Research - Zurich, 8803 R\"{u}schlikon, Switzerland}
\author{Anastasios Petropoulos}\affiliation{IBM Research - Zurich, 8803 R\"{u}schlikon, Switzerland}\affiliation{University of Patras, 26504 Rio Achaia, Greece}
\author{Theodore Antonakopoulos}\affiliation{University of Patras, 26504 Rio Achaia, Greece}
\author{Bipin Rajendran}\email{bipin@njit.edu}\affiliation{New Jersey Institute of Technology (NJIT), Newark, NJ 07102, USA}
\author{Abu Sebastian}\email{ase@zurich.ibm.com}\affiliation{IBM Research - Zurich, 8803 R\"{u}schlikon, Switzerland}
\author{Evangelos Eleftheriou}\affiliation{IBM Research - Zurich, 8803 R\"{u}schlikon, Switzerland}
\date{\today}

\begin{abstract}
Deep neural networks (DNNs) have revolutionized the field of artificial intelligence and have achieved unprecedented success in cognitive tasks
such as image and speech recognition. Training of large DNNs, however, is computationally intensive and this has motivated the search for novel
computing architectures targeting this application. A computational memory unit with nanoscale resistive memory devices organized in crossbar arrays
could store the synaptic weights in their conductance states and perform the expensive weighted summations in place in a non-von Neumann manner.
However, updating the conductance states in a reliable manner during the weight update process is a fundamental challenge that limits the training
accuracy of such an implementation. Here, we propose a mixed-precision architecture that combines a computational memory unit performing the weighted
summations and imprecise conductance updates with a digital processing unit that accumulates the weight updates in high precision. A combined
hardware/software training experiment of a multilayer perceptron based on the proposed architecture using a phase-change memory (PCM) array achieves
97.73\% test accuracy on the task of classifying handwritten digits (based on the MNIST dataset), within 0.6\% of the software baseline. The
architecture is further evaluated using accurate behavioral models of PCM on a wide class of networks, namely convolutional neural networks,
long-short-term-memory networks, and generative-adversarial networks. Accuracies comparable to those of floating-point implementations are achieved
without being constrained by the non-idealities associated with the PCM devices. A system-level study demonstrates 173$ \times $ improvement in energy
efficiency of the architecture when used for training a multilayer perceptron compared with a dedicated fully digital 32-bit implementation.
\end{abstract}
\maketitle

\section{Introduction}
 Loosely inspired by the adaptive parallel computing architecture of the brain, deep neural networks (DNNs) consist of layers of neurons and
weighted interconnections called synapses. These synaptic weights can be learned using known real-world examples to perform a given classification
task on new unknown data. Gradient descent based algorithms for training DNNs have been successful in achieving human-like accuracy in several
cognitive tasks. The training typically involves three stages. During forward propagation, training data is propagated through the DNN to determine
the network response. The final neuron layer responses are compared with the desired outputs to compute the resulting error. The objective of the
training process is to reduce this error by minimizing a cost function. During backward propagation, the error is propagated throughout the network
layers to determine the gradients of the cost function with respect to all the weights. During the weight update stage, the weights are updated based
on the gradient information. This sequence is repeated several times over the entire dataset, making training a computationally intensive
task\cite{Y2015lecunNature}. Furthermore, when training is performed on conventional von Neumann computing systems that store the large weight
matrices in off-chip memory, constant shuttling of data between memory and processor occurs. These aspects make the training of large DNNs very
time-consuming, in spite of the availability of high-performance computing resources such as general purpose graphical processing units (GPGPUs).
Also, the high-power consumption of this training approach is prohibitive for its widespread application in emerging domains such as the internet of
things and edge computing, motivating the search for new architectures for deep learning.

In-memory computing is a non-von Neumann concept that makes use of the physical attributes of memory devices organized in a computational memory unit
to perform computations in-place \cite{Y2018ielminiNatureElectronics}. Recent demonstrations include the execution of bulk bit-wise
operations\cite{Y2016seshadriArXiv}, detection of temporal correlations \cite{Y2017sebastianNatComm}, and  matrix-vector
multiplications\cite{Y2016huDAC,Y2017sheridanNatNano,legalloTED2018, Y2018legalloNatureElectronics,Y2017burrAPX,li2018analogue}. The matrix-vector
multiplications can be performed in constant computational time complexity using crossbar arrays of resistive memory (memristive) devices
\cite{Y2015wongNatureNano,Y2016huDAC,Y2018ielminiNatureElectronics}. If the network weights are stored as the conductance states of the memristive devices at the crosspoints,
then the weighted summations (or matrix-vector multiplications) necessary during the data-propagation stages (forward and backward) of training DNNs
can be performed in-place using the computational memory, with significantly reduced data movement \cite{Y2015burrTED,prezioso2015,Y2016gokmenFN,li2018,yao2017,Sun2018}. However, realizing accurate and gradual modulations of the conductance of the memristive devices for the weight update stage has posed a major challenge in utilizing computational memory to achieve accurate DNN training \cite{Yu2016}.

The conductance modifications based on atomic rearrangement in nanoscale memristive devices are stochastic, nonlinear, and asymmetric as well as of limited granularity
\cite{Wouters2015,Nandakumar2018}. This has led to significantly reduced classification accuracies compared with software baselines in training
experiments using existing memristive devices\cite{Y2015burrTED}. 
There have been several proposals to improve the precision of synaptic devices. A multi-memristive architecture uses multiple devices per synapse and programs one of them chosen based on a global selector during weight update\cite{Boybat2018}. Another approach, which uses multiple devices per synapse, further improves the precision by assigning significance to the devices as in a positional number system such as base two. The smaller updates are accumulated in the least significant synaptic device and periodically carried over to higher significant analog memory devices accurately\cite{Agarwal2017}. Hence, all devices in the array must be reprogrammed every time the carry is performed, which brings additional time and energy overheads. A similar approach uses a 3T1C (3 transistor 1 capacitor) cell to accumulate smaller updates and transfer them to PCM periodically using closed-loop iterative programming\cite{Y2018ambrogioNature}. So far, the precision offered by these more complex and expensive synaptic architectures has only been sufficient to demonstrate software-equivalent accuracies in end-to-end training of multi-layer perceptrons for MNIST image classification. All these approaches use a parallel weight update scheme by sending overlapping pulses from the rows and columns,  thereby implementing an approximate outer product and potentially updating all the devices in the array in parallel. Each outer product needs to be applied to the arrays one at a time (either after every training example or one by one after a batch of examples), leading to a large number of pulses applied to the devices. This has significant ramifications for device endurance, and the requirements on the number of conductance states to achieve accurate training\cite{Y2016gokmenFN,Yu2016}. Hence, this weight update scheme is best suited for fully-connected networks trained one sample at a time and is limited to training with stochastic gradient descent without momentum, which is a severe constraint on its applicability to a wide range of DNNs. 
The use of convolution layers, weight updates based on a mini-batch of samples as opposed to a single one, optimizers such as ADAM\cite{Kingma2014}, and techniques
such as batch normalization\cite{ioffe2015} have been crucial for achieving high learning accuracy in recent DNNs.

Meanwhile, there is a significant body of work in the conventional digital domain using reduced precision arithmetic for accelerating DNN training\cite{Y2015guptaICML,Y2015courbariauxANIPS,merolla2016,Zhang2017,hubara2017}. Recent studies show that it is possible to reduce the precision of the weights used in the multiply-accumulate operations (during the forward and backward propagations) to even 1 bit, as long as the weight gradients are accumulated in high-precision \cite{Y2015courbariauxANIPS}. This indicates the possibility of accelerating DNN training using programmable low-precision computational memory, provided that we address the challenge of reliably maintaining the high-precision gradient information. Designing the optimizer in the digital domain rather than in the analog domain permits the implementation of complex learning schemes that can be supported by general-purpose computing systems, as well as maintaining the high-precision in the gradient accumulation, which is necessary to be as high as 32-bit for training state-of-the-art networks \cite{Micikevicius2017}. However, in contrast to the fully digital mixed-precision architectures which uses statistically accurate rounding operations to convert high-precision weights to low precision weights and subsequently make use of error-free digital computation, the weight updates in analog memory devices using programming pulses are highly inaccurate and stochastic. Moreover, the weights stored in the computational memory are affected by noise and temporal conductance variations. Hence, it is not evident if the digital mixed-precision approach translates successfully to a computational memory based deep learning architecture.

Building on these insights, we present a mixed-precision computational memory architecture (MCA) to train DNNs. First, we experimentally demonstrate
the efficacy of the architecture to deliver performance close to equivalent floating-point simulations on the task of classifying handwritten digits from the MNIST dataset. Subsequently, we validate the approach
through simulations to train a convolutional neural network (CNN) on the CIFAR-10 dataset, a long-short-term-memory (LSTM) network on the Penn
Treebank dataset, and a generative-adversarial network (GAN) to generate MNIST digits.


\section{Results}
\subsection{Mixed-precision computational memory architecture}\label{sec:mp}
\begin{figure}[h!]
\includegraphics[width=\textwidth]{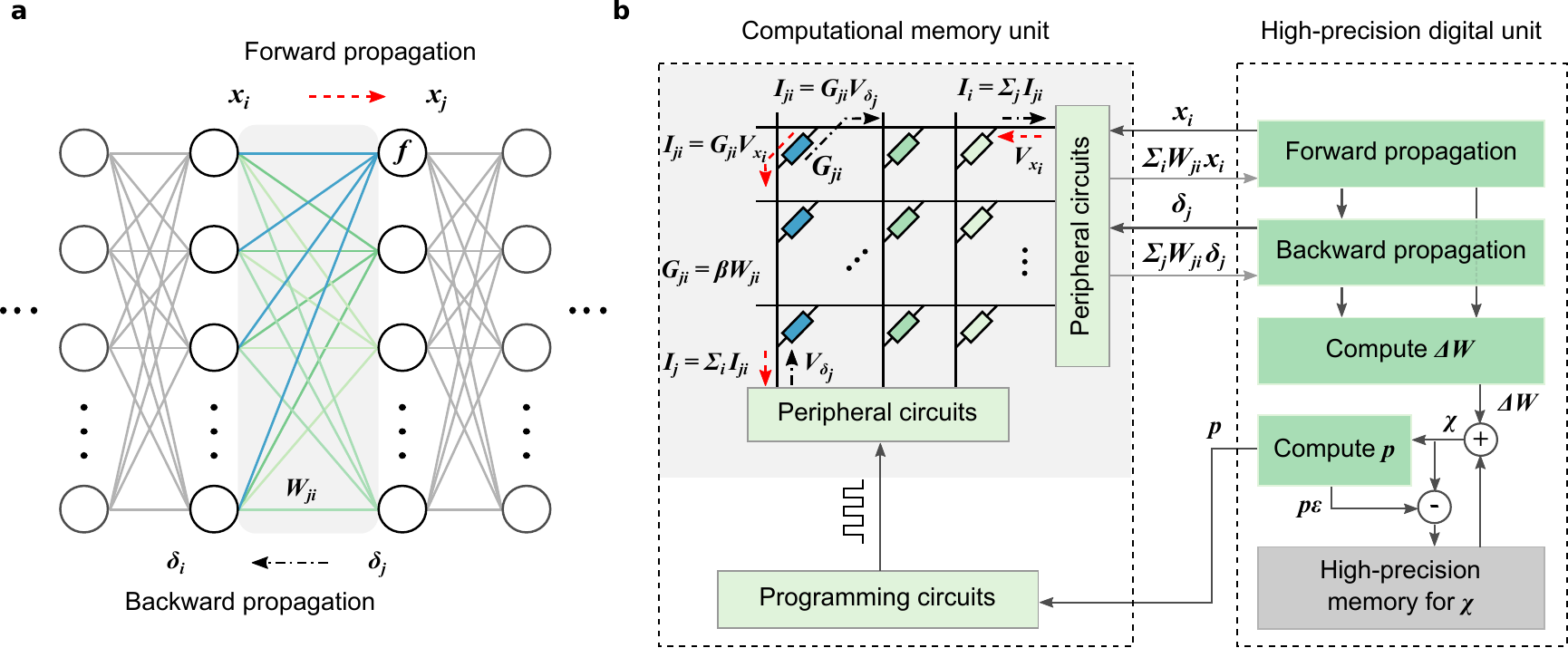}
\caption{\label{fig:mpa}\textbf{Mixed-precision computational memory architecture for deep learning.} \textbf{a} A neural network consisting of
layers of neurons with weighted interconnects. During forward propagation, the neuron response, $x_i$, is weighted according to the connection
strengths, $W_{ji}$, and summed. Subsequently, a non-linear function, $f$, is applied to determine the next neuron layer response, $x_j$. During
backward propagation, the error, $\delta_j$, is back-propagated though the weight layer connections, $W_{ji}$, to determine the error, $\delta_i$, of
the preceding layer.  \textbf{b} The mixed-precision architecture consisting of a computational memory unit and a high-precision digital unit. The
computational memory unit has several crossbar arrays whose device conductance values $G_{ji}$ represent the weights $W_{ji}$ of the DNN layers. The
crossbar arrays perform the weighted summations during the forward and backward propagations. The resulting $x$ and $\delta$ values are used to
determine the weight updates, $\Delta W$, in the digital unit. The $\Delta W$ values are accumulated in the variable, $\chi$. The conductance values
are updated using $p=\floor{\chi/\epsilon}$ number of pulses applied to the corresponding devices in the computational memory unit, where $\epsilon$
represents the device update granularity.}
\end{figure}

A schematic illustration of the MCA for training DNNs is shown in Fig.~\ref{fig:mpa}. It consists of a computational memory unit comprising several
memristive crossbar arrays, and a high-precision digital computing unit. If the weights $W_{ji} $ in any layer of a DNN
(Fig.~\ref{fig:mpa}\textbf{a}) are mapped to the device conductance values $G_{ji}$ in the computational memory with an optional scaling factor, then
the desired weighted summation operation during the data-propagation stages of DNN training can be implemented as follows. For the forward
propagation, the neuron activations, $x_{i}$, are converted to voltages, $V_{x_{i}}$, and applied to the crossbar rows. Currents will flow through
individual devices based on their conductance and the total current through any column, $I_j= \Sigma _i G_{ji}V_{x_{i}}$,  will correspond to $\Sigma
_i W_{ji}x_i$, that becomes the input for the next neuron layer. Similarly, for the backward propagation through the same layer, the voltages
$V_{\delta_j}$ corresponding to the error $\delta_j$ are applied to the columns of the same crossbar array and the weighted sum obtained along the
rows, $\Sigma _j W_{ji}\delta_j$, can be used to determine the error $\delta_i$ of the preceding layer.

The desired weight updates are determined as $\Delta W_{ji} = \eta \delta_j x_i$, where $\eta$ is the learning rate. We accumulate these updates in a
variable $\chi$ in the high-precision digital unit. The accumulated weight updates are transferred to the devices by applying single-shot programming
pulses, without using an iterative read-verify scheme. Let $\epsilon$ denote the average conductance change that can be reliably programmed into the
devices in the computation memory unit using a given pulse. Then, the number of programming pulses $p$ to be applied can be determined by rounding
$\chi/\epsilon$ towards zero. The programming pulses are chosen to increase or decrease the device conductance depending on the sign of $p$, and
$\chi$ is decremented by $p\epsilon$ after programming. Effectively, we are transferring the accumulated weight update to the device when it becomes
comparable to the device programming granularity. Note that the conductances are updated by applying programming pulses blindly without correcting
for the difference between the desired and observed conductance change. In spite of this, the achievable accuracy of DNNs trained with MCA is
extremely robust to the nonlinearity, stochasticity, and asymmetry of conductance changes originating from existing nanoscale memristive devices
\cite{Y2018nandakumarISCAS}.

\subsection{Characterization and modeling of PCM devices}\label{sec:model}
\begin{figure}[h!]
\includegraphics[width=\textwidth]{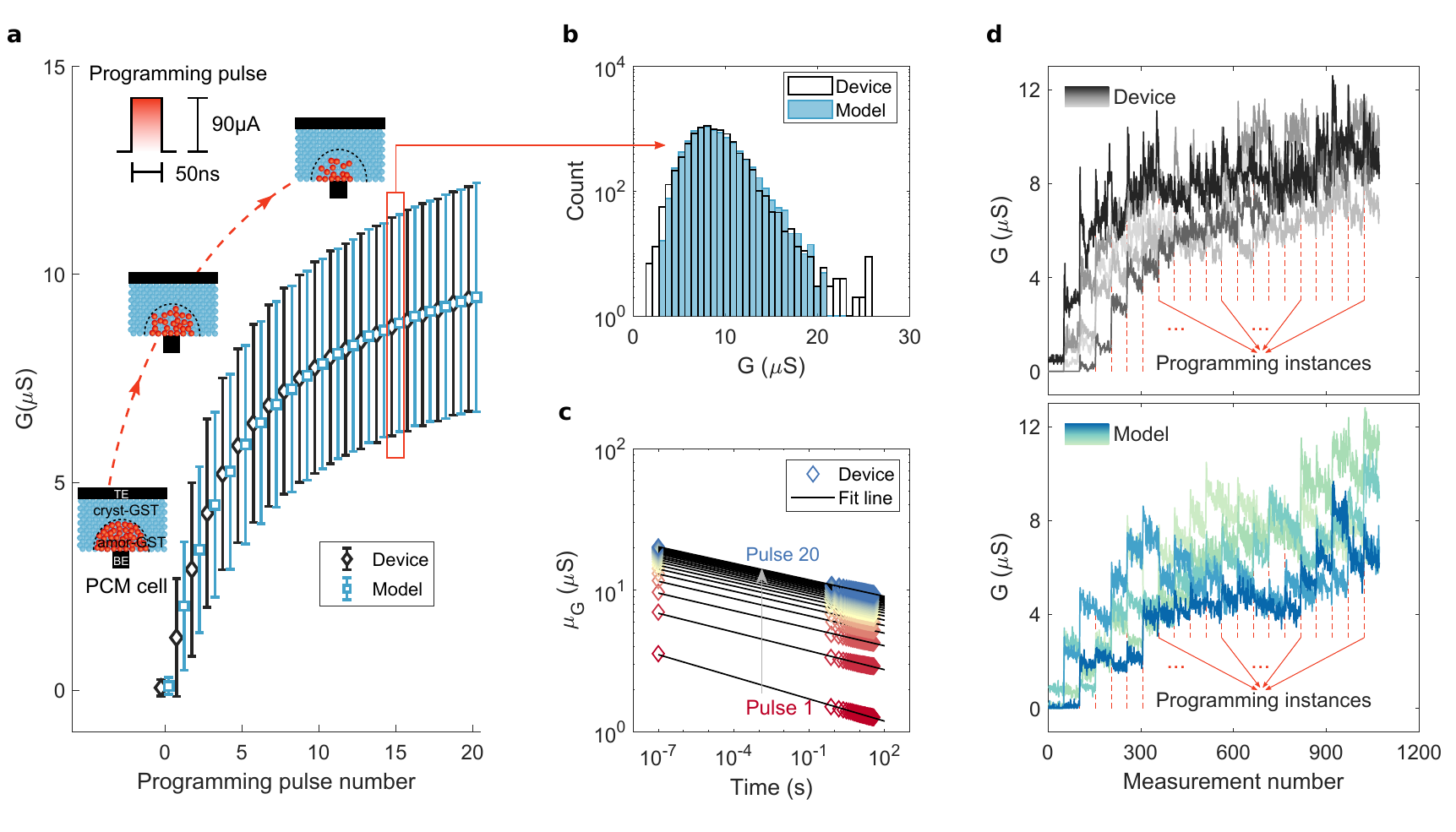}
\caption{\label{fig:PCM}\textbf{Phase-change memory characterization experiments and model response.} \textbf{a} The mean and standard deviation of
device conductance values (and the corresponding model response) as a function of the number of SET pulses of \unit[90]{$\mu$A} amplitude and
\unit[50]{ns} duration. The 10,000 PCM devices used for the measurement were initialized to a distribution around $0.06\,\mu$S. \textbf{b} The
distribution of conductance values compared to that predicted by the model after the application of 15 SET pulses. \textbf{c} The average conductance
drift of the states programmed after each SET pulse. The corresponding model fit is based on the relation, $G(t) = G(t_0)(t/t_0)^{-\nu}$, that
relates the conductance $G$ after time $t$ from programming to the conductance measurement at time $t_0$ and drift exponent $\nu$. \textbf{d}
Experimentally measured conductance evolution from 5 devices upon application of successive SET pulses compared to that predicted by the model. These
measurements are based on 50 reads that follow each of the 20 programming instances.}
\end{figure}

Phase-change memory (PCM) devices are used to realize the computational memory for the experimental validation of MCA. PCM is arguably the most
advanced memristive technology that has found applications in the space of storage-class memory\cite{Y2016burrJETCAS} and novel computing paradigms
such as neuromorphic computing \cite{Y2011kuzumNanoletters,Y2016tumaNatNano,Y2018sebastianJAP} and computational memory
\cite{Y2013cassinerioAdvMat,Y2017sebastianNatComm,legalloTED2018}. A PCM device consists of a nanometric volume of a chalcogenide phase-change
alloy sandwiched between two electrodes. The phase-change material is in the crystalline phase in an as-fabricated device. By applying a current
pulse of sufficient amplitude (typically referred to as the RESET pulse) an amorphous region around the narrow bottom electrode is created via
melt-quench process. The resulting ``mushroom-type'' phase configuration is schematically shown in Fig.~\ref{fig:PCM}\textbf{a}. The device will be in a high
resistance state if the amorphous region blocks the conductance path between the two electrodes. This amorphous region can be partially crystallized
by a SET pulse that heats the device (via Joule heating) to its crystallization temperature regime\cite{Y2014sebastianNatComm}. With the
successive application of such SET pulses, there is a progressive increase in the device conductance. This analog storage capability and the
accumulative behavior arising from the crystallization dynamics are central to the application of PCM in training DNNs.

We employ a prototype chip fabricated in \unit[90]{nm} CMOS technology integrating an array of doped Ge$_2$Sb$_2$Te$_5$ (GST) PCM devices (see methods).
To characterize the gradual conductance evolution in PCM, 10,000 devices are initialized to a distribution around \unit[0.06]{$\mu$S} and are
programmed with  a sequence of 20 SET pulses of amplitude \unit[90]{$\mu$A} and duration \unit[50]{ns}. The conductance changes show significant
randomness, which is attributed to the inherent stochasticity associated with the crystallization process \cite{Y2016legalloESSDERC}, together with
device-to-device variability \cite{Boybat2018,Y2016tumaNatNano}. The statistics of  cumulative conductance evolution are shown in Fig.~\ref{fig:PCM}\textbf{a}. The conductance evolves
in a state-dependent manner and tends to saturate with the number of programming pulses, hence exhibiting a nonlinear accumulative behavior. We analyzed the
conductance evolution due to the SET pulses, and developed a comprehensive statistical model capturing the accumulative
behavior that shows remarkable agreement with the measured data\cite{Y2018nandakumarJAP} (See Fig.~\ref{fig:PCM}\textbf{a},\textbf{b} and
Supplementary Note 1). Note that, the conductance response curve is unidirectional and hence asymmetric as we cannot achieve a progressive decrease
in the conductance values with the application of successive RESET pulses of the same amplitude.

The devices also exhibit a drift behavior attributed to the structural relaxation of the melt-quenched amorphous phase \cite{Y2018legalloAEM}. The
mean conductance evolution after each programming event as a function of time is plotted in Fig.~\ref{fig:PCM}\textbf{c}. Surprisingly, we find that
the drift re-initiates every time a SET pulse is applied\cite{Y2018nandakumarJAP} (see Supplementary Note 1), which could be attributed to the creation of a new unstable glass
state due to the atomic rearrangement that is triggered by the application of each SET pulse. In addition to the conductance drift, there are also
significant fluctuations in the conductance values (read noise) mostly arising from the $1/f$ noise exhibited by amorphous phase-change materials
\cite{Y2009nardonePRB}. The statistical model response from a few instances incorporating the programming non-linearity, stochasticity, drift, and
instantaneous read noise along with actual device measurements are shown in Fig.~\ref{fig:PCM}\textbf{d}, indicating the similar trend in conductance evolution
between the model and the experiment at an individual device level.

\subsection{Training experiment for handwritten digit classification}\label{sec:exp}
\begin{figure}[h!]
\includegraphics[width=\textwidth]{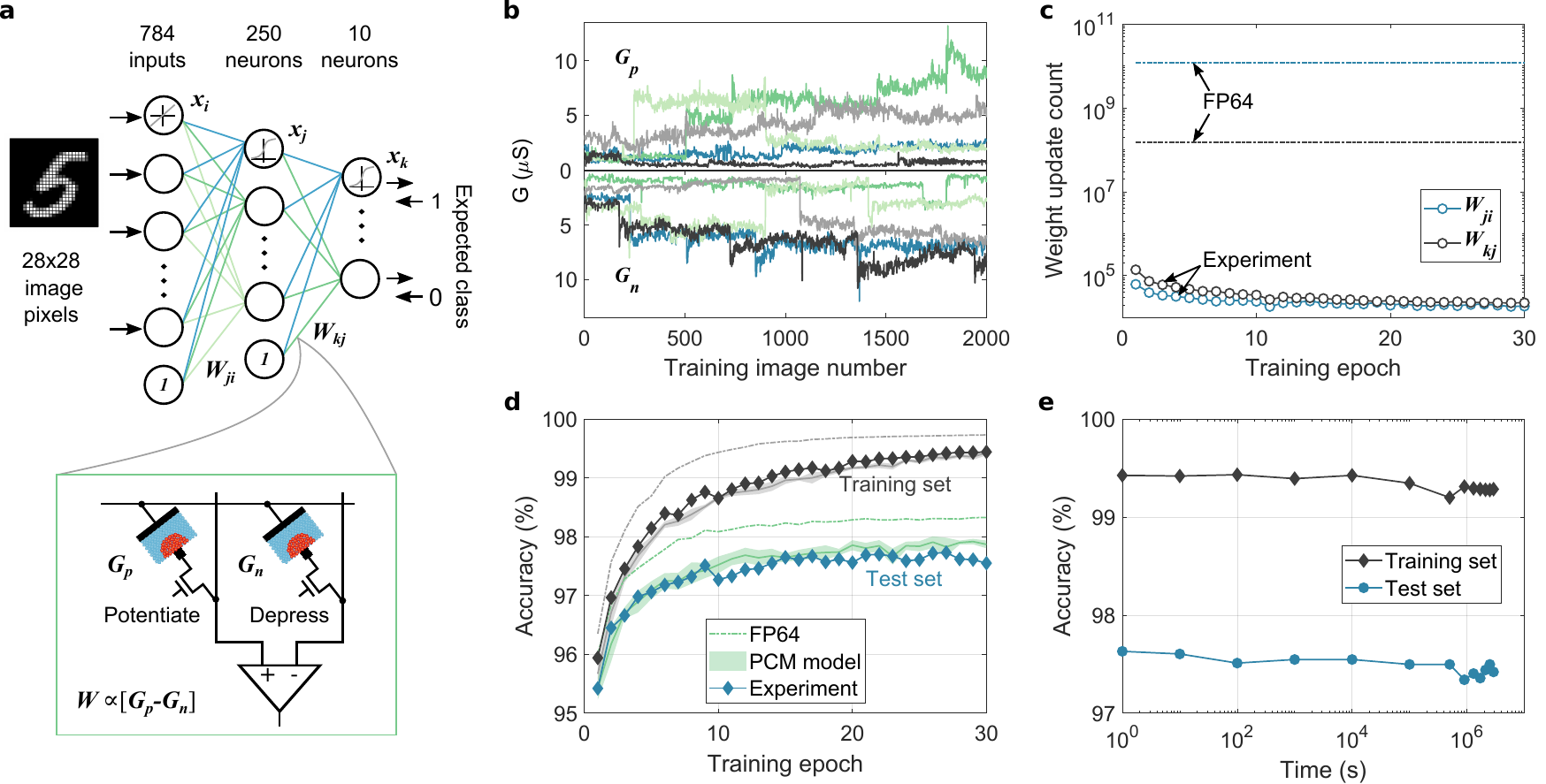}
\caption{\label{fig:exp}\textbf{MCA training experiment using on-chip PCM devices for handwritten digit classification.} \textbf{a} Network structure
used for the on-chip mixed-precision training experiment for MNIST data classification. Each weight, $W$,  in the network is realized as the
difference in conductance values of two PCM devices, $G_p$ and $G_n$. \textbf{b} Stochastic conductance evolution during training of $G_p$ and $G_n$
values corresponding to 5 arbitrarily chosen synaptic weights from the second layer. \textbf{c} The number of device updates per epoch from the two
weight layers in mixed-precision training experiment and high-precision software training (FP64), showing the highly sparse nature of weight update
in MCA. \textbf{d} Classification accuracies on the training and test set from the mixed-precision training experiment. The maximum experimental test
set accuracy, 97.73\%, is within 0.57\% of that obtained in the FP64 training . The experimental behavior is closely matched by the training
simulation using the PCM model. \textbf{e} Inference performed using the trained PCM weights on-chip on the training and test dataset as a function
of time elapsed after training showing negligible accuracy drop over a period of one month.}
\end{figure}

We experimentally demonstrate the efficacy of the MCA by training a two-layer perceptron to perform handwritten digit classification
(Fig.~\ref{fig:exp}\textbf{a}). Each weight of the network, $W$,  is realized using two PCM devices in a differential configuration ($W \propto
(G_p-G_n)$).  The 198,760 weights in the network are mapped to 397,520 PCM devices in the hardware platform (see methods). The network is trained
using 60,000 training images from the MNIST dataset for 30 epochs. The devices are initialized to a conductance distribution with mean  $1.6\,\mu$S
and standard deviation of  $0.83\,\mu$S. These device conductance values are read from hardware, scaled to the network weights, and used for the
data-propagation stages. The resulting weight updates are accumulated in the variable $\chi$. When the magnitude of $\chi$ exceeds $\epsilon$ (=
0.096, corresponding to an average conductance change of $0.77\,\mu$S per programming pulse), a $50\,$ns pulse with an amplitude of $90\,\mu$A is
applied to $G_p$ to increase the weight if $\chi>0$ or to $G_n$ to decrease the weight if  $\chi<0$;  $|\chi|$ is then reduced by $\epsilon$. These
device updates are performed using blind single-shot pulses without a read-verify operation and the device states are not used to determine the
number or shape of programming pulses. Since the continuous SET programming could cause some of the devices to saturate during training, a weight
refresh operation is performed every 100 training images to detect and reprogram the saturated synapses. After each training example involving a device update, all the devices in the second layer and 785 pairs of devices from the first layer are read along with the updated conductance values to use for the subsequent data-propagation step (see Methods). A separate validation experiment confirms that near identical results are obtained when the read voltage is varied in accordance with the neuron activations and error vectors for every matrix-vector multiplication during training and testing (see Supplementary Note 2). The resulting evolution of conductance pairs, $G_p$ and $G_n$, for
five arbitrarily chosen synapses from the second layer is shown in Fig.~\ref{fig:exp}\textbf{b}. It illustrates the stochastic conductance update,
drift between multiple training images, and the read noise experienced by the neural network during training. Also, due to the accumulate-and-program
nature of the mixed-precision training, only a few devices are updated after each image. In Fig.~\ref{fig:exp}\textbf{c}, the number of weight
updates per epoch in each layer during training is shown. Compared to the high-precision training where all the weights are updated after each image,
there are more than three orders of magnitude reduction in the number of updates in the mixed-precision scheme, thereby reducing the device programming
overhead.

At the end of each training epoch, all the PCM conductance values are read from the array and are used to evaluate the classification performance of the network  on the entire training set and on a disjoint set of 10,000 test images
(Fig.~\ref{fig:exp}\textbf{d}). The network achieved a maximum test accuracy of  $97.73\%$, only $0.57\%$ lower than the equivalent
classification accuracy of $98.30\%$ achieved in the high-precision training. The high-precision comparable training performance achieved by the MCA,
where the computational memory comprises noisy non-linear devices with highly stochastic behavior, demonstrates the existence of a solution to these
complex deep learning problems in the device-generated weight space. And even more remarkably, it highlights the ability of the MCA to successfully
find such solutions. We used the PCM model to validate the training experiment using simulations and the resulting training and test accuracies
are plotted in Fig.~\ref{fig:exp}\textbf{d}. The model was able to predict the experimental classification accuracies on both the training and the
test sets within \unit[0.3]{\%}, making it a valuable tool to evaluate the trainability of PCM-based computational memory for more complex deep
learning applications. The model was also able to predict the distribution of synaptic weights across the two layers remarkably well (see
Supplementary Note 3). It also indicated that the accuracy drop from the high-precision baseline training observed in the experiment is mostly attributed to PCM programming stochasticity (see Supplementary Note 4).
After training, the network weights in the PCM array were read repeatedly over time and the classification performance
(inference) was evaluated (Fig.~\ref{fig:exp}\textbf{e}). It can be seen that the classification accuracy drops by a mere \unit[0.3]{\%} over a time
period exceeding a month. This clearly illustrates the feasibility of using trained PCM based computational memory as an inference engine (see Methods).

The use of PCM devices in a differential configuration necessitates the refresh operation. Even though experiments show that the training methodology
is robust to a range of refresh schemes (see Supplementary Note 5), it does lead to additional complexity. But remarkably, the mixed-precision scheme
can deal with even highly asymmetric conductance responses such as in the case where a single PCM device is used to represent the synaptic weights.
We performed such an experiment realizing potentiation via SET pulses while depression was achieved using a RESET pulse. To achieve a
bipolar weight distribution, a reference conductance level was introduced (see Methods). By using different values of $\epsilon$ for potentiation and
depression, a maximum test accuracy of $97.47\%$ was achieved within 30 training epochs (see Supplementary Figure 1). These results
conclusively show the efficacy of the mixed-precision training approach and provide a pathway to overcome the stringent requirements on the device
update precision and symmetry hitherto thought to be necessary to achieve high performance from memristive device based learning
systems\cite{Y2015burrTED, Y2016gokmenFN, Kim2017, Y2018ambrogioNature}.

\subsection{Training simulations of larger networks}\label{sec:sim}
\begin{figure}[h!]
\includegraphics[scale =1]{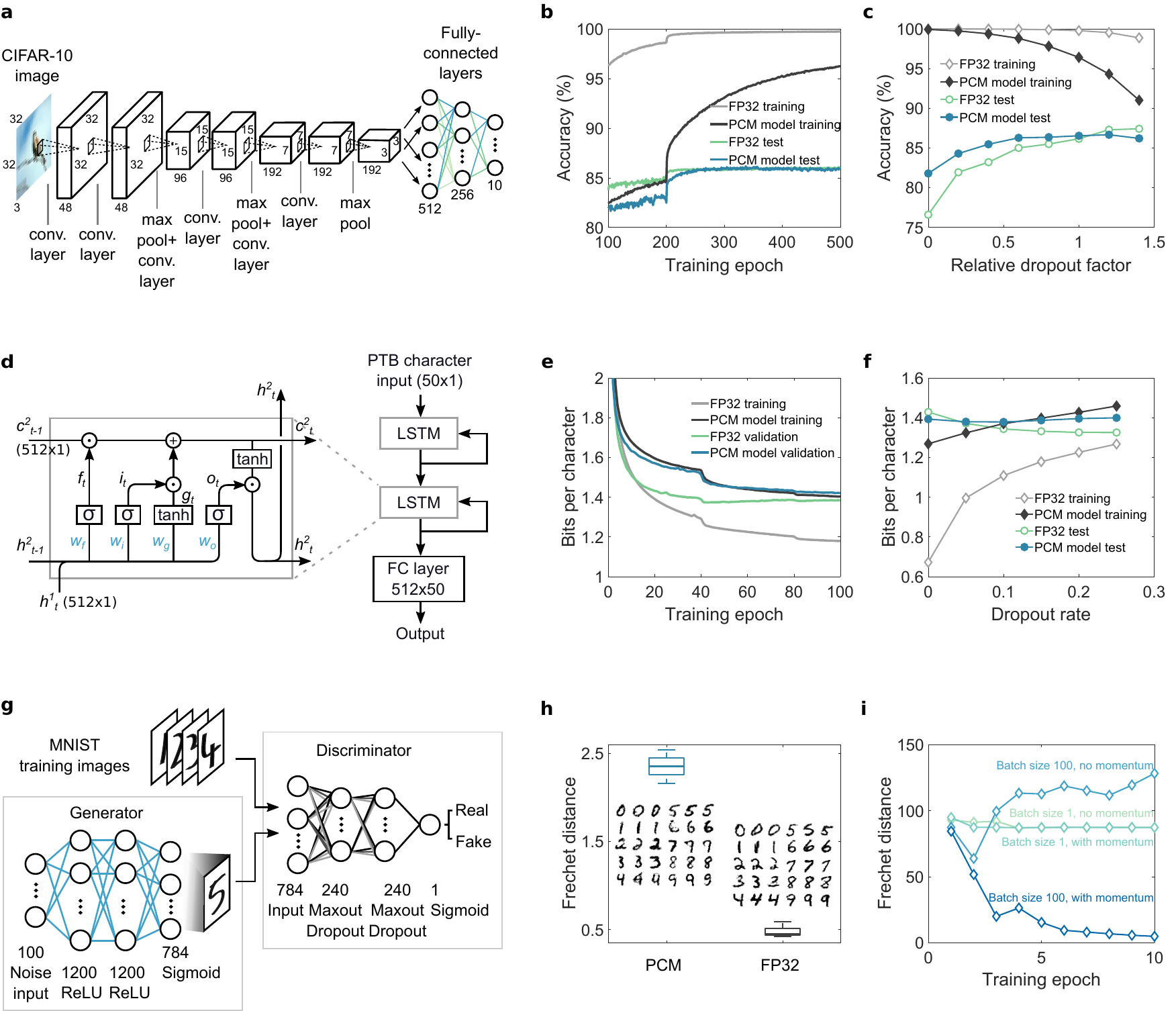}
\caption{\label{fig:sim}\textbf{MCA training validation on complex networks.} \textbf{a} Convolutional neural network for image classification on
CIFAR-10 dataset used for MCA training simulations. The two convolution layers followed by maxpooling and dropout are repeated thrice, and are
followed by three fully-connected layers. \textbf{b} The classification performance on the training and the test datasets during training. It can be
seen that the test accuracy corresponding to MCA-based training eventually exceeds that from the high-precision (FP32) training. \textbf{c} The
maximal training and test accuracies obtained as a function of the dropout rates. In the absence of dropout, MCA-based training significantly
outperforms FP32-based training. \textbf{d} The LSTM network used for MCA training simulations. Two LSTM cells with 512 hidden units followed by a
fully-connected (FC) layer are used. \textbf{e} The BPC as a function of the epoch number on training and validation sets shows that after 100
epochs, the validation BPC is comparable between the MCA and FP32 approaches. The network uses a dropout rate of 0.15 between non-recurring
connections. \textbf{f} The best BPC obtained after training as a function of the dropout rate indicates that without any dropout, MCA-based
training delivers better test performance (lower BPC) than FP32 training. \textbf{g} The GAN network used for MCA training simulations. The
generator and discriminator networks are fully-connected. The discriminator and the generator are trained intermittently using real images from the
MNIST dataset and the generated images from the generator. \textbf{h} The Frechet distance obtained from MCA training and the generated images are
compared to that obtained from the FP32 training. \textbf{i} The performance measured in terms of the Frechet distance as a function of the number of
epochs obtained for different mini-batch sizes and optimizers for FP32 training. To obtain convergence, mini-batch size greater than 1 and the
use of momentum are necessary in the training of the GAN.}
\end{figure}

The applicability of the MCA training approach to a wider class of  problems is verified by performing simulation studies based on the PCM model
described in Section \ref{sec:model}. The simulator is implemented as an extension to the TensorFlow deep learning framework. Custom TensorFlow
operations are implemented that take into account the various attributes of the PCM devices such as stochastic and nonlinear conductance update, read
noise, and conductance drift as well as the characteristics of the data converters (see Methods and Supplementary Note 6). This simulator is used to
evaluate the efficacy of the MCA on three networks: a CNN for classifying CIFAR-10 dataset images, an LSTM network  for character level language
modeling, and a GAN for image synthesis based on the MNIST dataset.

CNNs have become a central tool for many domains in computer vision and also for other applications such as audio analysis in the frequency domain.
Their power stems from the translation-invariant weight sharing that significantly reduces the number of parameters needed to extract relevant
features from images. As shown in Fig.~\ref{fig:sim}\textbf{a}, the investigated network consists of three sets of two convolution layers with ReLU
(rectified linear unit) activation followed by max-pooling and dropout, and three fully-connected layers at the end. This network has approximately
1.5 million trainable parameters in total (see methods and Supplementary Note 7). The convolution layer weights are mapped to the PCM devices of the
computational memory crossbar arrays by unrolling the filter weights and stacking them to form a 2D matrix \cite{Gokmen2017}. The network is trained
on the CIFAR-10 benchmark dataset using stochastic gradient descent (SGD) with a minibatch of 51 images and light data augmentation. The training and test classification accuracies as a function of training epochs are shown in
Fig.~\ref{fig:sim}\textbf{b}. The maximal test accuracy of the network trained via MCA ($ 86.46 \pm 0.25 \%$) is similar to that obtained from
equivalent high-precision training using 32-bit floating-point precision (FP32) ($ 86.24 \pm 0.19 \%$). However, this is achieved while having a significantly lower
training accuracy for MCA, which is suggestive of some beneficial regularization effects arising from the use of stochastic PCM devices to represent synaptic
weights. To understand this regularization effect further, we investigated the maximal training and test accuracies  as a function of the dropout
rates (see Fig.~\ref{fig:sim}\textbf{c})(see methods for more details). It was found that the optimal dropout rate for the network trained via MCA is
lower than that for the network trained in FP32. Indeed, if the dropout rate is tuned properly, the test accuracy of the network trained in FP32
could marginally outperform that of the one trained with MCA. Without any dropout, the MCA-trained network outperforms the one trained via FP32 which
suffers from significant overfitting.

LSTM networks are a class of recurrent neural networks used mainly for language modeling and temporal sequence learning. The LSTM cells are a natural
fit for crossbar arrays, as they basically consist of fully-connected layers \cite{li2019}. We use a popular benchmark dataset called Penn Treebank (PTB)\cite{penntreebank} for training the LSTM network (Fig.~\ref{fig:sim}\textbf{d}) using MCA. The network 
consists of two LSTM modules, stacked with a final fully-connected layer, and has a total of 3.3 million trainable parameters (see Supplementary Note 7).  The network is trained using sequences of text from the PTB dataset to
predict the next character in the sequence. The network response is a probability distribution for the next character in the sequence. 
The performance in a character level language modeling task is commonly measured using bits-per-character (BPC), which is a measure of how well the model is able to predict samples from the true underlying probability distribution of the dataset. A lower BPC corresponds to a better model.
The MCA based training and validation curves are shown in Fig.~\ref{fig:sim}\textbf{e}
(see methods for details). While the validation BPC from MCA and FP32 at the end of training are comparable, the difference between training
and validation BPC is significantly smaller in MCA. The BPC obtained on the test set after MCA and FP32 training for different dropout rates
are shown in Fig.~\ref{fig:sim}\textbf{f}. The optimal dropout rate that gives the lowest BPC for MCA is found to be lower than that for FP32,
indicating regularization effects similar to those observed in the case of the CNN.


GANs are neural networks trained using an recently proposed adversarial training method\cite{Goodfellow_2014_GAN}. The investigated network
has two parts: a generator network that receives random noise as input and attempts to replicate the distribution of the training dataset and a
discriminator network that attempts to distinguish between the training (real) images and the generated (fake) images. The network is deemed
converged when the discriminator is no longer able to distinguish between the real and the fake images. Using the MNIST dataset, we successfully
trained the GAN network shown in Fig.~\ref{fig:sim}\textbf{g} with MCA (see methods for details). The performance of the generator to replicate the
training dataset distribution is often evaluated using the Frechet distance (FD)\cite{Y2018_GAN_FD}. Even though the FD achieved by MCA training is
slightly higher than that of FP32 training, the resulting generated images appear quite similar (see Fig.~\ref{fig:sim}\textbf{h}).  The training of
GANs is particularly sensitive to the mini-batch size and the choice of optimizers, even when training in FP32. As shown in
Fig.~\ref{fig:sim}\textbf{i}, the solution converges to an optimal value only in the case of a mini-batch size of 100 and when  stochastic gradient
descent with momentum is used; we observed that the  solution  diverges in the other cases. Compared to alternate in-memory computing approaches
where both the propagation and weight updates are performed in the analog domain \cite{Y2018ambrogioNature}, a significant advantage of the proposed
MCA approach is its ability to seamlessly incorporate these more sophisticated optimizers as well as the use of mini-batch sizes larger than one
during training.

\section{Discussion}
We demonstrated via experiments and simulations that the MCA can train PCM based analog synapses in DNNs to achieve accuracies comparable to
those from the floating-point software training baselines. Here, we assess the overall system efficiency of the MCA for an exemplary problem and
discuss pathways for achieving performance superiority as a general deep learning accelerator. We designed an application specific integrated circuit
(ASIC) in \unit[14]{nm} low power plus (14LPP) technology to perform the digital computations in the MCA and estimated the training energy per image,
including that spent in the computational memory and the associated peripheral circuits (designed in 14LPP as well). The implementation was designed for the two-layer
perceptron performing MNIST handwritten digit classification used in the experiments of Section \ref{sec:exp}. For reference, an equivalent
high-precision ASIC training engine was designed in 14LPP using 32-bit fixed-point format for data and computations with an effective throughput of
 43k images/s at \unit[0.62]{W} power consumption (see Methods and Supplementary Note 8 for details).  In both designs, all the memory necessary for training was implemented with on-chip
static random-access memory. The MCA design resulted in 271$ \times $ improvement in energy consumption for the forward and backward stages of training.
Since the high-precision weight update computation and accumulation are the primary bottleneck for computational efficiency in the MCA, we
implemented the outer-products for weight update computation using low-precision versions of the neuron activations and back-propagated errors
\cite{hubara2017,Zhang2017,wu2018}, achieving comparable test accuracy with respect to the experiment of Section \ref{sec:exp} (see Supplementary
Note 8). Activation and error vectors were represented using signed 3-bit numbers and shared scaling factors. The resulting weight update matrices
were sparse with less than 1\% non-zero entries on average. This proportionally reduced accesses to the 32-bit $\chi$ memory allocated for accumulating the
weight updates. Necessary scaling operations for the non-zero entries were implemented using bit-shifts \cite{lin2015,wu2018}, thereby reducing the
computing time and hardware complexity. Additional device programming overhead was negligible, since on average only 1 PCM device was
programmed every 2 training images. This resulted in 139$ \times $ improvement in the energy consumption for the weight update stage in MCA with respect to the
32-bit design. Combining the three stages, the MCA achieved a 11.5$ \times $ higher throughput at 173$ \times $ lower energy consumption with respect to the
32-bit implementation (see Table~\ref{table_ener_sum}). We also compared the MCA with a fully digital mixed-precision ASIC in 14LPP, which uses 4-bit weights and 8-bit activations/errors for the data propagation stages. This design uses the same weight update implementation as the MCA design, but replaces the computational memory by a digital multiply-accumulate unit. The MCA achieves an overall energy efficiency gain of 23$ \times $ with respect to this digital mixed-precision design (see Methods).

\begin{table}[ht]
	\setlength\extrarowheight{4pt}
	\centering
	\caption{Energy and time estimated based on application specific integrated circuit (ASIC) designs for processing one training image in MCA and corresponding fully digital 32-bit and mixed-precision designs. The numbers are for a specific two-layer perceptron with 785 input neurons, 250 hidden neurons, and 10 output neurons.} 
	\label{table_ener_sum}
	\begin{tabular}{|p{5cm}|p{2cm}|p{2cm}p{2cm}p{2cm}|p{2cm}|}
		\hline
		Architecture & Parameter  & Forward propagation & Backward propagation & Weight update & Total \\ \hline \hline
		\multirow{2}{*}{32-bit design}& Energy  & 5.62\,$ \mu  $J & 0.09\,$ \mu  $J & 8.64\,$ \mu  $J & 14.35\,$ \mu $J \\
		& Time & 7.31\,$ \mu $s & 0.59\,$ \mu $s & 15.36\,$ \mu $s & 23.27\,$ \mu $s \\ \hline 
		{Fully digital mixed-precision design}& Energy  & 1.78\,$ \mu  $J & 0.016\,$ \mu  $J & 0.076\,$ \mu  $J & 1.87\,$ \mu $J \\
		{4-bit weights, 8-bit activations/errors}& Time & 6.41\,$ \mu $s & 0.13\,$ \mu $s & 0.79\,$ \mu $s & 7.33\,$ \mu $s \\ \hline 
		\multirow{2}{*}{MCA -- computational memory} & Energy  & 7.27\,nJ & 2.15\,nJ & 0.05\,nJ & \\
		& Time  & 0.26\,$ \mu  $s&  0.13\,$ \mu  $s & -- &  \\ \hline
		\multirow{2}{*}{MCA -- digital unit} & Energy  & 8.91\,nJ & 2.73\,nJ & 61.98\,nJ &\\
		& Time & 0.34\,$ \mu $s & 0.09\,$ \mu $s & 1.19\,$ \mu $s & \\ \hline
		\multirow{2}{*}{MCA -- total} & Energy & 16.18\,nJ & 4.88\,nJ & 62.03\,nJ & 83.08\,nJ\\
		& Time & 0.61\,$ \mu $s &0.22\,$ \mu $s &1.19\,$ \mu $s & 2.01\,$ \mu $s\\ \hline
	\end{tabular}
\end{table}

While the above study was limited to a simple two-layer perceptron, deep learning with MCA could generally have the following benefits over fully
digital implementations. For larger networks, digital deep learning accelerators as well as the MCA will have to rely on dynamic random-access memory (DRAM) to store the model parameters, activations, and errors, which will significantly increase the cost of access to those variables compared with on-chip SRAM. Hence, implementing DNN weights in nanoscale memory devices could enable large neural networks to be fit on-chip without expensive
off-chip communication during the data propagations. Analog crossbar arrays implementing matrix-vector multiplications in $ \mathcal{O}(1) $ time
permit orders of magnitude computational acceleration of the data-propagation stages\cite{merrikh2017,li2018analogue,Y2016gokmenFN}. Analog in-memory
processing is a desirable trade-off of numerical precision for computational energy efficiency, as a growing number of DNN architectures are being
demonstrated to support low precision weights\cite{Y2015courbariauxANIPS, Y2018KailashArxiv}. In contrast to digital mixed-precision ASIC
implementations, where the same resources are shared among all the computations, the dedicated weight layers in MCA permit  more efficient inter and
intra layer pipelines\cite{shafiee2016,Y2017_PipeLayer}. Handling the control of such pipelines for training various network topologies adequately with optimized array-to-array communication, which is a non-trivial task, will be crucial in harnessing the efficiency of the MCA for deep learning. Compared with the fully analog accelerators being explored\cite{Agarwal2017, Y2018ambrogioNature, Kim2017}, the MCA requires an additional high-precision digital memory of same size as the model, and the need to access that memory during the weight update stage. 
However, the digital implementation of the optimizer in the MCA provides high-precision gradient accumulation and the flexibility to
realize a wide class of optimization algorithms, which are highly desirable in a general deep learning accelerator. 
Moreover, the MCA significantly relaxes the analog synaptic device requirements, particularly those related to linearity, variability, and update precision to realize high-performance learning machines. 
In contrast to the periodic carry approach\cite{Agarwal2017, Y2018ambrogioNature}, it avoids the need of reprogramming all the weights at specific intervals during training. Instead, single-shot blind pulses are applied to chosen synapses at every weight update, resulting in sparse device programming. This relaxes the overall reliability and endurance requirements of the nanoscale devices and reduces the time and energy spent to program them.

In summary, we proposed a mixed-precision computational memory architecture for training DNNs and experimentally demonstrated its ability to deliver
performance close to equivalent 64-bit floating-point simulations. We used a prototype phase-change memory (PCM) chip to perform the training of a
two-layer perceptron containing 198,760 synapses on the MNIST dataset. We further validated the approach by training a CNN on the CIFAR-10 dataset, an LSTM network on the Penn Treebank dataset, and a GAN to generate MNIST digits. The training of these larger networks was performed through simulations using a PCM
behavioral model that matches the characteristics of our prototype array, and achieved accuracy close to 32-bit software training in all the three cases.
 We also showed evidence for inherent regularization effects originating
from the non-linear and stochastic behavior of these devices that is indicative of futuristic learning machines exploiting rather than overcoming the
underlying operating characteristics of nanoscale devices. These results show that the proposed architecture can be used to
train a wide range of DNNs in a reliable and flexible manner with existing memristive devices, offering a pathway towards more energy-efficient deep
learning than with general-purpose computing systems.

\clearpage
\section*{Methods}
{\footnotesize
	
\subsection*{PCM-based hardware platform}
The experimental hardware platform is built around a prototype phase-change memory (PCM) chip that contains 3 million PCM
devices\cite{Y2010closeIEDM}. The PCM devices are based on doped Ge$_2$Sb$_2$Te$_5$ (GST) and are integrated into the  chip in \unit[90]{nm} CMOS
baseline technology. In addition to the PCM devices, the  chip integrates the circuitry for device addressing, on-chip ADC for device readout, and
voltage- or current-mode device programming. The experimental platform comprises a high-performance analog-front-end (AFE) board that contains a
number of digital-to-analog converters (DACs) along with discrete electronics such as power supplies, voltage and current reference sources. It also
comprises an FPGA board that implements the data acquisition and the digital logic to interface with the PCM device under test and with all the
electronics of the AFE board. The FPGA board also contains an embedded processor and ethernet connection that implement the overall system control
and data management as well as the interface with the host computer. The embedded microcode allows the execution of the multi-device programming and
readout experiments that implement the matrix-vector multiplications and weight updates on the PCM chip. The hardware modules implement the interface
with the external DACs used to provide various voltages to the chip for programming and readout, as well as the interface with the memory device
under test, i.e., the addressing interface, the programming-mode or read-mode interfaces, etc.

The PCM device array is organized as a matrix of 512 word lines (WL) and 2048 bit lines (BL). Each individual device along with its access transistor
occupies an area of \unit[50]{F$^2$} (F is the technology feature size, \unit[$\text{F}=90$]{nm}). The PCM devices were integrated into the chip in
\unit[90]{nm} CMOS technology using a sub-lithographic key-hole transfer process\cite{breitwischVLSI2007}. The bottom electrode has a radius of
$\sim$ \unit[20]{nm} and a length of $\sim$ \unit[65]{nm}. The phase change material is $\sim 100$ nm thick and extends to the top electrode, whose
radius is $\sim 100$ nm. The selection of one PCM device is done by serially addressing a WL and a BL. The addresses are decoded and they then drive
the WL driver and the BL multiplexer. The single selected device can be programmed by forcing a current through the BL with a voltage-controlled
current source.  For reading a PCM device, the selected BL is biased to a constant voltage of 300 mV by a voltage regulator via a voltage
$V_\mathrm{read}$ generated off-chip. The sensed current, $I_\mathrm{read}$, is integrated by a capacitor, and the resulting voltage is then
digitized by the on-chip 8-bit cyclic ADC. The total time of one read is \unit[$1$]{$\mu$s}. The readout characteristic is calibrated via the use of
on-chip reference polysilicon resistors. For programming a PCM device, a voltage $V_\mathrm{prog}$ generated off-chip is converted on-chip into a
programming current, $I_{\mathrm{prog}}$. This current is then mirrored into the selected BL for the desired duration of the programming pulse. Each
programming pulse is a box-type rectangular pulse with duration of \unit[10]{ns} to \unit[400]{ns} and amplitude varying between 0 and 500\,$\mu$A.
The access-device gate voltage (WL voltage) is kept high at \unit[2.75]{V} during programming.  Iterative programming, which is used for device
initialization in our experiments,  is achieved by applying a sequence of programming pulses\cite{Y2011papandreouISCAS}. After each programming
pulse, a verify step is performed and the value of the device conductance programmed in the preceding iteration is read at a voltage of
\unit[0.2]{V}. The programming current applied to the PCM device in the subsequent iteration is adapted according to the sign of the value of the
error between the target level and read value of the device conductance. The programming sequence ends when the error between the target conductance
and the programmed conductance of the device is smaller than a desired margin or when the maximum number of iterations (20) has been reached. The
total time of one program-and-verify step is approximately \unit[$2.5$]{$\mu$s}.

\subsection*{Mixed-precision training experiment}

\textbf{Network details:} The network used in the experiment had 784 inputs and 1 bias at the input layer, 250 sigmoid neurons and 1 bias at the hidden layer, and 10 sigmoid neurons at the output layer. The network was trained by minimizing the mean square error loss function with stochastic gradient descent (SGD). The network was trained with a batch size of 1, meaning that the weight updates were computed after every training example. We used a fixed learning rate of 0.4. We used the full MNIST training dataset of 60,000 images for training the network, and the test dataset of 10,000 images for computing the test accuracy. The order of the images was randomized for each training epoch. Apart from normalizing the gray-scale images, no additional pre-processing was performed on the training and test sets. 

\textbf{Differential PCM experiment:} 397,520 devices were used from the PCM hardware platform to represent the two-layer network
(Fig.~\ref{fig:exp}\textbf{a})  weights in a differential configuration. The devices were initialized to a conductance distribution with mean of $1.6\,\mu$S
and standard deviation of  $0.83\,\mu$S via iterative programming.  Since the platform allowed only serial access to the devices, all the conductance values were read from hardware and reported to the software that performed the forward and backward propagations. The weight updates were computed and accumulated in the $\chi$ memory in software. When the magnitude of $\chi$ exceeded $\epsilon$ for a particular weight, a $50\,$ns pulse with an amplitude of $90\,\mu$A was applied to the corresponding device ($G_p$ or $G_n$, depending on the sign of $\chi$) of the PCM chip.
The synaptic conductance to weight conversion was performed by a linear mapping between [-8$\,\mu$S, 8$\,\mu$S] in the conductance domain
and [-1, 1] in the weight domain. 

The PCM devices exhibit temporal variations in the conductance values such as conductance drift and read noise. As a result, each matrix
multiplication in every layer will see a slightly different weight matrix even in the absence of any weight update. However, the cost of re-reading
the entire conductance array for every matrix-vector multiplication in our experiment was prohibitively large due to the serial interface. Therefore, after each programming event to the PCM array, we read the conductance values of a subset of all the PCM devices along with the programmed device. Specifically, we read all the devices in the second layer and a set of 785 pairs of devices from the first layer in a round robin fashion after every device programming event. This approach faithfully captures the effects of PCM hardware noise and drift in the network propagations during training. For the weight refresh operation, the conductance pairs
in software were verified every 100 training examples and if one of the device conductance values was above \unit[8]{$\mu$S} and if their difference was less than
\unit[6]{$\mu$S}, both the devices were RESET using 500\,ns, 360\,$\mu$A pulses and their difference was converted to a number of SET pulses based on
an average observed conductance change per pulse. During this weight refresh, the maximum number of pulses was limited to 3 and the pulses were
applied to $G_p$ or $G_n$ depending on the sign of their initial conductance difference.

\textbf{Inference after training:} After training, all the PCM conductance values realizing the weights of the two-layer perceptron are read at different time intervals and used to evaluate the classification accuracy on the 60,000 MNIST training images and 10,000 test images. Despite the conductance drift, there was only negligible accuracy drop over a month. The following factors might be contributing to this drift tolerance. During on-chip training, different devices are programmed at different times and are drifting during the training process. The training algorithm could compensate for the error created by the conductance drift and could eventually generate a more drift resilient solution. Furthermore, the perceptron weights are implemented using the conductance difference of two PCM devices. This differential configuration partially compensates the effect of drift \cite{Y2018BoybatNVMTS}. Also, from the empirical relation for the conductance drift, we have $\frac{dG}{dt} \approx G(t_0)\times t_0^\nu\frac{(-\nu)}{t} \propto \frac{1}{t} $, which means that the conductance decay over time decreases as we advance in time. Drift compensation strategies such as using a global scaling factor\cite{legalloTED2018} could also be used in more complex deep learning models to maintain the accuracy over extended periods of time.

\textbf{Non-differential PCM experiment:} A non-differential PCM configuration for the synapse was tested in the mixed-precision training
architecture to implement both weight increment and decrement by programming the same device in either direction and hence avoid the conductance
saturation and the associated weight refresh overhead. The non-accumulative RESET behavior of the PCM device makes its conductance potentiation and
depression highly asymmetric and hence this experiment also validates tolerance of the mixed-precision training architecture to such programming
asymmetry. We conducted the same training experiment as before for the MNIST digit classification, except that now each synapse was realized using a
single PCM with a reference level to realize bipolar weights.  The experiment requires 198,760 PCM devices. Potentiation is implemented by
$90\,\mu$A, $50\,$ns SET pulses and depression is implemented using $400\,\mu$A, $50\,$ns RESET pulses. In the mixed-precision architecture,
this asymmetric conductance update behavior is compensated for by using different $\epsilon$s for potentiation and depression. We used $\epsilon_P$
corresponding to $0.77\,\mu$S for weight increment and $\epsilon_D$ corresponding to $8\,\mu$S for weight decrement.

Ideally, the reference levels could be fabricated using any resistive device which is one time programmed to the necessary conductance level.  In the
case of PCM devices, due to their conductance drift, it is more suitable to implement the reference levels using the PCM technology which follows the same
average drift behavior of the devices that represent the synapses. In this experiment, we used the average conductance of all the PCM devices read
from the array to represent the reference conductance level ($G_{ref}$), and hence the network weights $W \propto (G-G_{ref})$. For the experiment, the
devices are initialized to a distribution with mean conductance of  $4.5\,\mu$S   and standard deviation of $1.25\,\mu$S. While a narrower
distribution was desirable\cite{Glorot10a}, it was difficult to achieve in the chosen initialization range. However, this was compensated by mapping
the conductance values to a narrower weight range at the beginning of the training. This weight range is progressively relaxed over the next few
epochs. The conductance range [0.1, 8]$\,\mu$S is mapped to [-0.7, 0.7] during epoch 1. Thereafter, the range is incremented in uniform steps/epoch
to [-1, 1] by epoch 3 and is held constant thereafter. Also, the mean value of the initial conductance distribution was chosen to be slightly higher
than the midpoint of the conductance range to compensate for the conductance drift.
Irrespective of the conductance change asymmetry, the training in MCA achieves a maximum  training accuracy of 98.77\% and a maximum test accuracy of 97.47\%  in 30 epochs (see Supplementary Figure 1).

\subsection*{Training simulations of larger networks with PCM model}
\textbf{Simulator:} The simulator is implemented as an extension to the Tensorflow deep learning framework. The Tensorflow operations that can be
implemented on computational memory are replaced with custom implementations. For example, the Ohm's law and Kirchhoff's circuit laws replace the
matrix-vector multiplications and are implemented based on the PCM model described in Section \ref{sec:model}. The various non-idealities associated
with the PCM devices such as limited conductance range, read noise and conductance drift as well as the quantization effects arising from the data converters (with 8-bit
quantization) are incorporated while performing the matrix-vector multiplications. The synaptic weight update routines are also replaced with custom
ones that implement the mixed-precision weight update accumulation and stochastic conductance update using the model of the PCM described
in Section \ref{sec:model}. In the matrix-vector multiplication operations, the weight matrix represented using stochastic PCM devices is multiplied by 8-bit fixed-point neuron activations or normalized 8-bit fixed-point error vectors. Since the analog current accumulation along the wires in the computational memory unit can be assumed to have arbitrary precision, the matrix-vector multiplication between the noisy modeled PCM weights and quantized inputs is computed in 32-bit floating-point. In order to model the peripheral analog to digital conversion, the matrix-vector multiplication results are quantized back to 8-bit fixed-point. We evaluated the effect of ADC precision on DNN training performance based on the MNIST handwritten digit classification problem\cite{Y2018nandakumarISCAS}.  The accuracy loss due to quantization was estimated to be approximately 0.1\% in MCA for 8-bit ADC, which progressively increases with reduced precision. It is possible to reduce the ADC precision down to 4-bit at the cost of a few percentage drop in accuracy for reduced circuit complexity. However, we chose to maintain an 8-bit ADC precision for the training of networks to maintain comparable accuracies with the 32-bit precision baseline. The remaining training operations such as activations, dropout, pooling, and weight update computations use 32-bit floating-point precision.

Hence, these operations use original Tensorflow implementations in the current simulations and can be performed in the digital unit of the MCA in its eventual hardware implementation.
Additional details on the simulator can be found in Supplementary Note 6.

\textbf{CNN:} The CNN used for the simulation study has 9 layers   -  6 convolution layers and 3 fully-connected layers \cite{parneetblog}. A pooling
layer is inserted after every two convolution layers. Together, the network has approximately 1.5 million parameters. ReLU activation is used at all convolution and fully-connected layers and softmax activation is used at the output layer. Dropout regularization
technique is employed after the pooling layers and in-between the fully-connected layers. Light data augmentation is also employed in the form of
random image flipping and random adjustments of brightness and contrast. To map the convolution kernels to crossbar arrays, the filters are stretched
out to 1D arrays and horizontally stacked on the memristive crossbar \cite{Gokmen2017}. The different image patches are extracted from the input
image, stretched out and finally rearranged to form the columns of a large matrix. The convolution can now be computed by performing the
matrix-matrix multiplication between these two matrices and subsequently, reordering the output. In other words, the forward propagation of neuron
activations and the backward propagation of the errors can be implemented as matrix-vector multiplications. The weight update for a single image patch can be computed as the
outer product of activation and error vectors, and the weight updates arising from all image patches are averaged to obtain the total weight update corresponding to the entire input image. During the extraction process of the image patches, the original image is padded with zero pixels at
the border, to ensure that the output of the convolution layer has the same image size as the input.  Considering an input image of size $n\times n$
and $d$ channels and $m$ convolution kernels of size $k\times k$, the dimensions of the input matrix is $dk^2\times n^2$ and the dimensions of the
matrix on the crossbar array is $dk^2\times m$. The convolution operation can therefore be performed in $n^2$ matrix-vector multiplication cycles.
The pooling operations of the CNN are performed in the conventional digital domain. The weight updates computed from
all the weight layers are accumulated in $\chi$ and are subsequently transferred to the modeled PCM devices organized in crossbar arrays. SGD with cross-entropy loss function is used for training (for additional details see Supplementary Note 7).

\textbf{Testing of the regularization effect observed during training of the CNN:} We observed that MCA achieves higher test accuracy with lower
training accuracy compared to the FP32 training. This is a desirable effect referred to as regularization and techniques such as dropout are
typically employed to achieve this. We suspected that the stochastic nature of the synaptic device prevents an over-fitting in this architecture and
hence allows to generalize better. To test this hypothesis, we ran both FP32 and MCA training simulations using varying dropout factors while keeping
the other hyperparameters to be the same. We scale both dropout rates ($0.5$ between the fully-connected and $0.25$ after the pooling layers) with a
scaling factor which takes the values $0.0,0.2,0.4,0.8,1.0,1.2,$ and $1.4$. The resulting maximal test accuracies are depicted in
Fig.~\ref{fig:sim}\textbf{c}. 
As dropout rates are reduced, MCA training achieves higher test accuracies compared with FP32 training, indicating the inherent regularization achieved via MCA.

\textbf{LSTM:} The LSTM network trained using MCA for the task of character-level language modelling contains roughly 3.3 million parameters
(Fig.~\ref{fig:sim}\textbf{d}). An LSTM cell takes as input a hidden state from the previous time step $h^{l}_{t-1}$ and the training data $x_t$ or
the hidden state from the previous layer $h^{l-1}_{t}$ and generates a new hidden state $h^l_t$ and updates a cell state $c^l$ using the weights
$w_f^l$, $w_i^l$, $w_g^l$, $w_o^l$, for $l = 1, 2$ according to the following relations:
\begin{align}
\begin{pmatrix}i_t\\f_t\\o_t\\g_t\end{pmatrix} &=\begin{pmatrix}\sigma\\\sigma\\\sigma\\tanh\end{pmatrix}\left[ W\begin{pmatrix} x_t \text{ or } h_t^{l-1}\\h_{t-1}^l\end{pmatrix} +b^l \right]\\ \\
c_t^l & = f_t \odot c_{t-1}^l + i_t\odot g_t \\
h_t^l & = o_t\odot tanh(c_t^l)
\end{align}
where $sigmoid$ ($\sigma$) and $tanh$ are applied element-wise and $\odot$ is the element-wise multiplication. $W$ is obtained by stacking $w_f^l$,
$w_i^l$, $w_g^l$, $w_o^l$. $i_t$, $f_t$, $o_t$, $g_t$, $h_t^l$, and $c_t^l$ are of dimension $n = 512$ and $x_t$ is of dimension $m = 50$. The weight
matrix, $W$, is of dimension $4n\times (n+m)$ in the first layer and $4n\times 2n$ in the second layer. $b^l$ is a $4n$-dimensional bias vector.
Dropout is applied to the non-recurrent connections (i.e. at the output of each LSTM cell) \cite{zaremba2014}. The output of the second LSTM cell is fed through a
fully-connected layer with 50 output neurons and then through a softmax activation unit. The weights in all the layers are represented and trained
using PCM device models organized in crossbar arrays (see Supplementary Note 7). The remaining gating operations are performed in the conventional
digital domain.

The PTB dataset has 5.1 million characters for training, 400k characters for validation, and 450k character for testing. The vocabulary contains 50
different characters and the data is fed into the network as one-hot encoded vectors of dimension ($50\times 1$), without any embedding. Each vector
has a single distinct location marked as 1 and the rest are all zeros. We used SGD with cross-entropy loss function for training. The training performance is evaluated using the bits-per-character (BPC) metric, which is the average cross-entropy loss evaluated with base 2 logarithm. 

\textbf{Testing the regularization effect during training of the LSTM:} To study the regularization effect, the LSTM network was trained with
different dropout rates, from 0.0 to 0.25 in steps of 0.05. The resulting minimal training and test BPC can be seen in
Fig.~\ref{fig:sim}\textbf{f}. The performance on the test dataset was less sensitive to the dropout rate in MCA and it achieved lower BPC
without any dropout compared to FP32 based training.  This result is also consistent with previously reported studies of training LSTMs using
memristive crossbars \cite{Y2018gokmen}. Considering the rather low number of parameters in the network, the test BPC results compare favorably with the current state-of-the-art methods \cite{merity2018}.

\textbf{GAN:} A multi-layer perceptron (MLP) implementation of GAN\cite{Goodfellow_2014_GAN} was employed with approximately 4 million trainable
parameters. The generator has 2 hidden layers with ReLU activations and an output layer of sigmoid activation. The discriminator has 2 hidden layers
with Maxout activations (maximum out of its five inputs) and a sigmoid output layer. Due to the large number of parameters in the discriminator, we
used dropout regularization with dropout rates 20\% and 30\% respectively after its first and second hidden layers. The GAN is trained using 50,000
images from the MNIST dataset and its performance is evaluated using 10,000 test images. $G_\text{LOSS}$ and $D_\text{LOSS}$, which represent the
loss functions minimized to train the generator and discriminator respectively, are as follows:
\begin{align}
G_\text{LOSS} & = \sum log(D(G(z)) \label{gloss}\\
D_\text{LOSS} & = \sum (log(D(x)) + log(1 - D(G(z)))) \label{dloss}
\end{align}

where, $x$ and $z$ respectively are the MNIST images and the random input noise. $G(.)$ and $D(.)$ represent the generator and the discriminator
networks and summation is over the training samples. Both the networks are trained with SGD using a learning rate of 0.1 and a momentum of 0.5. The weights in all the layers are represented and trained using PCM device models organized in crossbar arrays (see Supplementary Note 7 for more details on training).

The training performance of the GAN is evaluated using the Frechet distance (FD), a metric that is robust even in the presence of common failure
modes in  GAN\cite{GAN_large-scale_study, Y2017heuselArxiv, Y2018liuArxiv}. FD makes use of features extracted from a particular layer in a CNN
trained to classify the training dataset\cite{MNISTFeature}.
\begin{equation}\label{eqn:fd}
FD = ||\mu_{X} - \mu_{G}||_{2}^{2} + Tr(C_{X} + C_{G} - 2(C_{X} \times C_{G})^{1/2})
\end{equation}
$\mu_{X}$ and $\mu_{G}$ denote the mean values of features computed for the train/test dataset and the generated dataset, respectively. $C_{X}$ and
$C_{G}$ are covariance matrices of features computed for the train/test dataset and the generated dataset, respectively. $Tr$ is the trace operator
over a matrix and $||.||_{2}$ is the 2-norm operator.

\textbf{Study of batch size and optimizer for GAN:} To study the sensitivity of training GANs to the mini-batch size and the choice of optimizer used
for training, we trained FP32 (baseline) GAN implementation with two mini-batch sizes of 1 and 100 and with two different optimizer types: SGD without momentum and SGD with momentum while keeping all the other hyper-parameters the same. Fig.~\ref{fig:sim}\textbf{i}
compares results in these cases. Non-unity batch size with momentum was necessary to train the GAN successfully in our case. In all the other cases
the solution seems to diverge; the discriminator loss goes to zero and the generator loss diverges. Similar behavior is observed frequently in
generative networks\cite{Arjovsky2017}.

\subsection*{Energy estimation of MCA and comparison}
The energy efficiency of the MCA compared with a conventional 32-bit fixed-point digital design {and a digital mixed-precision design} for training was evaluated using respective ASIC implementations in 14LPP technology (see Supplementary Note 8). {The three} designs were customized  to perform the training operations of a two-layer perceptron trained to classify MNIST handwritten digits. They also accommodated all the necessary SRAM memory on-chip, avoiding the cost of off-chip memory access. The network had 784 inputs, 250 hidden neurons, and 10 output neurons as in the training experiment presented in Section \ref{sec:exp}. For  simplicity of the ASIC design, we used ReLU activations for the hidden layer neurons and L2SVM\cite{Tang2013DeepLU} for the error computation at the output layer. The network was trained using SGD with a batch size of 1.

Cycle-accurate register transfer level (RTL) models of the 32-bit all-digital design{, digital mixed-precision design,} and the digital unit of the MCA were developed. A testbench
infrastructure was then built to verify the correct behavior of the models using the Cadence NCsim simulator. Once the behavior was
verified, the RTL models were synthesized in Samsung 14LPP technology using Cadence Genus Synthesis Solution software. The synthesized netlists were then imported to Cadence Innovus software where the netlists were subjected to 
backend physical design steps of placing, clock tree synthesis, and routing. At the end of routing, the post route netlists were exported with which post route simulations were carried out in NCsim simulator. During each simulation, an activity file is generated which contain toggle count data for all the nets in the netlist. The activity data along with the parasitics extracted from the netlist were used to perform an accurate power estimation on the post route design for each stage of operation (forward propagation, backward propagation, weight update), with the Innovus software. For power estimation, a supply voltage of \unit[0.72]{V} was used in the designs while an operating frequency of \unit[500]{MHz} was used to clock  the {fully digital} designs and the digital unit of MCA.
Power numbers were then converted to energy by multiplying by respective time windows. The energy estimations from the digital unit were combined with those from the computational memory to determine the overall performance of
the MCA implementation. The computational memory unit of the MCA was designed separately with the necessary peripheral circuits to integrate it with
the digital unit of the MCA. Note that most of the memristive device technologies including PCM are amenable to back end of line (BEOL) integration, thus enabling their integration with mainstream front end CMOS technology. This allows the crossbar array peripheral circuits to be designed in 14LPP. To support the device programming, a separate power supply line might be necessary. Larger currents where necessary can be supported by fabricating wider or a parallel combination of transistors. Operating at 2\,GHz, the computational memory unit  used 16-bit wide bus for data transfer, pulse width modulators to apply digital variables as analog voltages to the input of the crossbar array, and ADCs to read the resulting output currents. The time and energy consumption of the peripheral circuits were obtained from circuit simulations in 14LPP technology. The energy consumption of the analog computation was estimated assuming the average device
conductance of 2.32\,$\mu$S observed from the hardware training experiment.   The average programming energy for the PCM conductance updates in the computational memory unit was estimated based on the SET and RESET (for weight refresh) pulse statistics from the training experiment. The device programming was executed in parallel to the weight update computation in the digital unit. In contrast to the hardware experiment, the weight refresh operation was distributed across training examples as opposed to periodically refreshing the whole array during training, which allowed it to be performed in parallel with the weight update computation in the digital unit (see Supplementary Note 8). 

The details on the design of the {three} training architectures and the corresponding energy estimations can be found in Supplementary Note 8. A summary of energy and computing time for the forward and backward data propagations and the weight update stages for {the} designs are listed in Table~\ref{table_ener_sum}. The energy and time are reported as average numbers for a single training example. Since the designs are operating at a different precision, throughput is reported as training examples processed per second. The baseline 32-bit implementation consumed 14.35\,$ \mu $J with a throughput of 43k images per second. Computational memory enabled an average energy gain of $271$ and an acceleration of $9.61$ for the forward and backward propagation stages with respect to the 32-bit design. The low-precision implementation of the outer product led to reduced precision multipliers (3-bit compared to 32-bit) and sparse access to the 32-bit $ \chi $ memory. These factors led to 139$ \times $ improvement in energy consumption for the digital weight update accumulation stage in MCA, with 2130 non-zero updates to the $ \chi $ memory being performed per training image. We verified via training simulations using the PCM model that the low-precision outer product optimization of the weight update could maintain comparable test accuracy as that from the PCM hardware training experiment (see Supplementary Note 8). Energy consumption due to the PCM programming in the computational memory was negligible compared with the energy spent in the digital unit of MCA for the weight update stage. Overall, the MCA consumed 83.08\,nJ per image and achieved 496k images/s throughput. The digital mixed-precision design followed a similar architecture as that of the MCA. However, the computational memory was replaced by a multiply-accumulate unit for the data propagation stages, optimized to use 4-bit weights and 8-bit activations or errors. Note that activations and error vectors have additional bit-shift based scaling factors to represent their actual magnitude. The digital mixed-precision design had the same reduced precision weight update scheme as the MCA, and hence a similar energy efficiency for weight updates. However, for the data propagation stages, the 4-bit weights were obtained from the 32-bit weights read from the on-chip SRAM. The requirement to read a high-precision memory for the data propagations is avoided in the in-memory computing architecture. As a result, MCA maintained an energy efficiency gain of 85$ \times $ during the data propagation stages, and 23$ \times $ overall with respect to the digital mixed-precision design. The fully digital mixed-precision design consumed $ 1.87\,\mu $J per training example with a throughput of 136k images per second.

\clearpage
\section*{References}
\bibliography{MPDL}
\section*{Acknowledgments}
We thank C. Malossi, N. Papandreou and C. Bekas for discussions, M. Braendli and M. Stanisavljevic for help with digital design tools, and our colleagues in IBM TJ Watson Research Center, in particular M. BrightSky, for
help with fabricating the PCM prototype chip used in this work. A.S. acknowledges funding from the European Research Council (ERC) under the European
Union's Horizon 2020 research and innovation programme (grant agreement number 682675).
\section*{Author Contributions}
S.R.N., M.L.G., I.B., A.S. and E.E. conceived the mixed-precision computational memory architecture for deep learning. S.R.N. developed the PCM model
and performed the hardware experiments with the support of M.L.G. and I.B.. C.P., V.J. and G.M. developed the TensorFlow simulator and performed the
training simulations of the large networks. G.K. designed the 32-bit digital training ASIC and the digital unit of the MCA, and performed their
energy estimations. R.K.A. designed the computational memory unit of the MCA and performed its energy estimation. U.E., A.P. and T.A. designed,
built, and developed the software of the hardware platform hosting the prototype PCM chip used in the experiments. S.R.N., M.L.G. and A.S. wrote the
manuscript with input from all authors. M.L.G., B.R. A.S. and E.E. supervised the project.
\section*{Conflict of Interest Statement}
S.R.N., M.L.G., C.P., V.J., G.M., I.B., G.K., R.K.A., U.E., A.P., A.S. and E.E. were employed by the company IBM Research - Zurich. The remaining authors declare that the research was conducted in the absence of any commercial or financial relationships that could be construed as a potential conflict of interest.

\end{document}